\documentclass[english]{article}
\usepackage[T1]{fontenc}
\usepackage[latin1]{inputenc}
\usepackage{geometry}
\usepackage{float}
\usepackage{mathrsfs}
\usepackage{amsmath}
\usepackage{amssymb}
\usepackage{graphicx}
\usepackage{subfigure}
\usepackage{caption}
\usepackage{array}
\usepackage{booktabs}
\usepackage{tabularx}
\usepackage{multirow}

\makeatletter

\floatstyle{ruled}
\newfloat{algorithm}{tbp}{loa}
\providecommand{\algorithmname}{Algorithm}
\floatname{algorithm}{\protect\algorithmname}

\usepackage{amsthm}

\usepackage{mathrsfs}

\usepackage{amsfonts}

\usepackage{epsfig}

\usepackage{bm}

\usepackage{mathrsfs}

\usepackage{enumerate}

\@ifundefined{definecolor}{\@ifundefined{definecolor}
	{\@ifundefined{definecolor}
		{\usepackage{color}}{}
	}{}
}{}

\usepackage{subfig}\usepackage[all]{xy}

\newcounter{hypA}

\newcounter{hypB}

\usepackage{babel}\date{}

\usepackage{babel}

\makeatother

\usepackage{babel}

\newcolumntype{Y}{>{\centering\arraybackslash}X}

\begin{document}
	
	%+Title
	\begin{center}
		
		{\Large \textbf{Bayesian Inference for Latent Chain Graphs}}
		
		\vspace{0.5cm}
		
		BY DENG LU$^{1}$, MARIA DE IORIO$^{2}$, AJAY JASRA$^{3}$ 
		\& GARY L. ROSNER$^{4}$
		
		{\footnotesize $^{1}$Department of Statistics \& Applied Probability,
			National University of Singapore, Singapore, 117546, SG.}
		{\footnotesize E-Mail:\,} \texttt{\emph{\footnotesize denglu@u.nus.edu.sg}}\\
		{\footnotesize $^{2}$Yale-NUS, Singapore, 138527, SG. \&}
		{\footnotesize Department of Statistical Science, University College London, UK}
		{\footnotesize E-Mail:\,} \texttt{\emph{\footnotesize maria@yale-nus.edu.sg}}\\
		{\footnotesize $^{3}$Computer, Electrical and Mathematical Sciences and Engineering Division,
			King Abdullah University of Science and Technology, Thuwal, 23955, KSA.}
		{\footnotesize E-Mail:\,} \texttt{\emph{\footnotesize ajay.jasra@kaust.edu.sa}}\\
		{\footnotesize $^{4}$Oncology Biostatistics and Bioinformatics, Sidney Kimmel Comprehensive Cancer Center, Johns Hopkins University, Baltimore, MD 21205, USA.}
		{\footnotesize E-Mail:\,} \texttt{\emph{\footnotesize grosner1@jhmi.edu}}
	\end{center}
	
	\begin{abstract}
		In this article we consider Bayesian inference for partially observed Andersson-Madigan-Perlman (AMP) Gaussian chain graph (CG) models.
		Such models are of particular interest in applications such as biological networks and financial time series. The model itself features a variety of constraints which make both prior modeling and computational inference challenging. We develop a framework for the aforementioned challenges, using a sequential Monte Carlo
		(SMC) method for statistical inference. Our approach is illustrated on both simulated data as well as real case studies from university graduation rates and a pharmacokinetics study. 
	\end{abstract}
	
	\section{Introduction}
	Two common  approaches to  probabilistically  describe conditional dependence structures are based on undirected graphs (Markov networks) and directed acyclic graphs (DAG) (Bayesian networks), e.g.~\cite{pearl}. 
	The vertices of the graph represent  variables, while the presence (absence) of an edge between two vertices indicates possible dependence (independence) between the two corresponding variables.  Applications of Markov networks include  biological networks, e.g., to study the dependence structure among genes from expression data \cite{Chun 2014,Dobra 2004} and financial time series for forecasting and predictive portfolio analysis \cite{Carvalho 2007,Wang 2011}; while Bayesian networks occur in expert system research for providing rapid absorption and propagation of evidence  \cite{lau88, pearl86}, path analysis for computing implied correlations \cite{werm80}, and psychometrics for causal pathways \cite{boere, kana}. 
	
	Chain graphs  \cite{lau84,laur,werm}  provide an elegant unifying point of view on Markov and Bayesian networks. 
	These models allow for edges that are both directed and undirected and do not contain any semi-directed
	cycles. 
	Although chain graphs were first introduced in the late eighties,  most research has focused on Bayesian networks and Gaussian Graphical Models. 
	Recently they are receiving more attention as they  have proved to be very useful in applications due to their ability to represent symmetric and non-symmetric relationships between the random variables of interest.  However,  there exist in the literature several different ways of interpreting chain graphs and what conditional independencies they encode, giving rise to different so-called \textit{chain graph interpretations}. Whilst a directed edge works as in DAGs when it comes to representing independence, the undirected edge can be understood in different ways, giving rise to the different interpretations. This implies that chain graphs can represent every independence model achievable by any DAG, whereas the opposite does not hold. The most common interpretations 
	are the Lauritzen-Wermuth-Frydenberg (LWF) interpretation, the Andersson-Madigan-Perlman interpretation and the multivariate regression (MVR) interpretation. Each interpretation has its own way of determining conditional independences in a CG and each interpretation subsumes another in terms of representable independence models; see \cite{sonn}.  Moreover, 
	\cite{lau02} discuss causal interpretation of chain graphs.
	
	This work focuses on  chain graphs of the AMP type \cite{and,levi01} and develops a statistical framework to infer the chain graph structure from data measured on a set of variables of interest with the goal of understanding their association pattern.  Many inferential procedures have concentrated  on the estimation of the parameters characterizing the graph, \textit{given the chain graph topology}.   For example,  \cite{drton} consider likelihood-based parameter inference for chain graph models that are directly observed. Nevertheless, in the machine learning literature, methods  based on optimization have been proposed to perform inference on the graph structure itself in the context of chain graphs, e.g.~\cite{mcc14,pena14}. We consider the case where
	not only are the parameters unknown, but the chain graph itself is unobserved. The main objective is to perform Bayesian inference in such a context.  
	
	There are several contributions that propose Bayesian methods in related contexts. In \cite{silva1}, the author proposes a method to perform full Bayesian inference for acyclic directed mixed graphs using
	Markov chain Monte Carlo (MCMC) methods. There are several similarities between the model there (based upon structural equation models (SEM) e.g .~\cite{bollen}) and the one developed here. However, the main difference
	is given by  the structure of the latent graph and the associated constraints in our case. These latter constraints are more stringent and lead to more complexity in the computational strategy. In \cite{wang} a similar model to the one presented in \cite{silva1} is developed, which uses a spike-and-slab prior to induce sparsity in the graph. The computational scheme is expected to outperform that in \cite{silva1} for a large number
	of nodes. The prior structure in \cite{wang} could in principle  be extended to accommodate chain graph structures  at the cost of introducing constraints that could make computations infeasible. 
	
	The contributions of this article are as follows. Based upon the likelihood method in \cite{drton} we develop a new Bayesian model for latent AMP chain graphs. This model can also incorporate covariate information, if available.
	We introduce a sequential Monte Carlo (SMC) method as in \cite{delm:06} that
	improves upon MCMC-based methods in this context. Our approach is applied to real case studies from university graduation rates and a pharmacokinetics study. We find the performance of our algorithm to be stable and robust.  
	
	This article is structured as follows. In Section \ref{sec:model} we describe our model and prior specifications. In Section \ref{sec:alg} we introduce the SMC algorithm. In Section \ref{sec:res} we present a simulation study and two real-life applications. In the  appendix we detail further elements of our algorithm in Section \ref{sec:alg}.
	
	\section{Model}\label{sec:model}
	
	\subsection{Likelihood}
	
	Let $Y=(Y_1,\ldots, Y_p)\in\mathbb{R}^p$ be a random vector whose  elements correspond to  $p\in\mathbb{N}$ nodes of a graph. We assume we have  $m\in\mathbb{N}$ observations on $Y$: $y_{1:m}$, $y_i\in\mathbb{R}^{p}$.
	
	A graph $G = (V,E)$ is described by a set of nodes $V$ and edges $E$, with variables $Y_1,\ldots,Y_p$ placed at the nodes. The edges define  the global conditional independence structure of the distribution. An AMP chain graph is a graph whose every edge is directed or undirected such that it does not contain any semi-directed cycles, that is, it contains no
	path from a node $v$ to itself with at least one directed edge such that all directed edges have the same orientation. Each graph can be identified by an adjacency matrix. Let   $A$ be a $p \times p$ matrix with entries $(a_{ij})_{1\leq i,j\leq p}$ where
	$a_{ij}\in\{0,\dots,r\}$ (let $r=3$, which is a notation used later on) with $a_{ii}=0$, for $i\neq j$
	$$
	a_{ij} = \left\{\begin{array}{ll}
	0 & \textrm{ no edge between}~i~\textrm{and}~j\\
	1 & \textrm{ undirected edge between}~i~\textrm{and}~j\\
	2 & \textrm{ directed edge from}~i~\textrm{to}~j \\
	3 &  \textrm{ directed edge from}~j~\textrm{to}~i 
	\end{array}\right.
	$$
	and, given the upper-triangular part of $A$, for $j<i$
	$$
	a_{ji} = \left\{\begin{array}{ll}
	a_{ij} & \textrm{if}~a_{ij}\in\{0,1\} \\
	2 &  \textrm{if}~a_{ij}=3 \\
	3 &  \textrm{if}~a_{ij}=2.
	\end{array}\right.
	$$
	Given $p$, a labelling of the nodes and the adjacency matrix $A$ define a  graph $G(A)$. Let $\mathcal{C}$ denote the set of possible chain graphs for a set of $p$ vertices.
	Given $A$, let  $\Omega$ be  a positive
	definite $p\times p$ (real-valued) matrix, such that if $a_{ij}\neq 1$ then $\omega_{ij}=0$. In addition, given $A$, for an arbitrary real-valued $p\times p$ matrix
	$B$, then, if $a_{ij}\neq 2$ $b_{ij}=0$. Finally set
	$$
	\Sigma(B,\Omega) = (I-B)^{-1} \Omega^{-1} (I-B')^{-1}
	$$
	where $I$ is the $p\times p$  identity matrix. To inherit the AMP chain graph property, $\Sigma(B,\Omega)$ should be a $p\times p$ positive definite matrix.
	
	The absence of semi-directed cycles implies that the vertex set of a chain graph can be partitioned into so-called chain components such that edges within a chain component are undirected whereas the edges between two chain components are directed and point in the same direction. More precisely, the vertex set of a chain graph $G(A)$ can be partitioned into subsets $\mathscr{T}(G(A))$
	such that all edges within each subset $\tau$ are un-directed and edges between two different subsets
	$\tau\neq \tau'$ are directed. In the following, we assume that the partition $\tau \in\mathscr{T}(G(A))$ is maximal, that is, any two
	vertices in a subset $\tau$  are connected by an un-directed path. 
	Recall that
	the parents of a set of nodes $\mathcal{X}$ of $G$ is the set $ \textrm{pa}
	( \mathcal{X} ) =\{ V_j \text{ s.t. }  
	V_j  \rightarrow V_i \in G,V_j\not\in  \mathcal{X} \quad \& \quad V_i \in\mathcal{X} \}$.
	Let $\textrm{pa}(\tau)$ be  parents of $\tau$, $B_{\tau}=(B_{ij})_{i\in\tau,j\in\textrm{pa}(\tau)}$ and $\Omega_{\tau}=(\omega_{ij})_{i,j\in\tau}$ sub-matrices of $B$ and $\Omega$, respectively. For a single observation $y\in\mathbb{R}^p$, $y_{\tau}\mid y_{\textrm{pa}(\tau)},B,\Omega\sim\mathcal{N}_{|\tau|}(B_{\tau}y_{\textrm{pa}(\tau)},\Omega_{\tau}^{-1})$, where, for $d\in\mathbb{N}$, 
	$\mathcal{N}_d(\mu,\Sigma)$ denotes the $d-$dimensional Gaussian distribution with  mean $\mu$ and covariance matrix $\Sigma$. Then   the joint density of $y\in\mathbb{R}^p$ given $B,\Omega,A$ is
	$$
	p(y\mid B,\Omega,A) = \prod_{\tau\in\mathscr{T}(G(A))}p(y_{\tau}\mid y_{\textrm{pa}(\tau)},B,\Omega).
	$$
	The likelihood is then given by 
	$$
	p(y_{1:m}\mid B,\Omega,A) = \prod_{i=1}^m p(y_i \mid B,\Omega,A)
	$$
	that is, the observations are i.i.d.~given $B,\Omega,A$. We remark that this structure is similar to the structural equation model  proposed by \cite{silva}, where also a Gaussian distribution is assumed for the variables corresponding to the nodes of the graph. Moreover, these above assumptions on the precision matrix ensure that the corresponding graph satisfies the AMP Markox property as discussed in \cite{and}.
	
	\subsection{Prior Distribution on the Chain Graph Space}
	
	We specify the prior on the set of possible chain graphs, by specifying a prior on the elements of the corresponding adjacency matrix. 
	For $1\leq i <j\leq p$, let 
	$$
	\mathbb{P}(a_{ij}=l\mid \pi_{ij}) = \pi_{ij}(l), \quad l = 0,\dots,r
	$$
	where $\pi_{ij}=(\pi_{ij}(0),\dots,\pi_{ij}(r) )$. We assume each vector  $\pi_{ij}$ to  follow a Dirichlet distribution with parameter $\alpha=(\alpha_0, \dots, \alpha_r )$ and the $\pi_{ij}$ to be conditionally independent given $\alpha$. Note that, if available, covariate information could be incorporated to model $\pi_{ij}$, as in \cite{tan} for example.
	Marginally, we have  
	\begin{align*}
		\mathbb{P}(a_{ij}= l \mid \alpha) &=\frac{1}{B(\alpha)} \int \prod_{k=0}^{r} \pi_{ij}(k)^{ \mathbb{I}(a_{ij}=l) } \prod_{k=0}^{r} \pi_{ij}(k)^{ \alpha_s-1 } d\pi_{ij}(0)... d\pi_{ij}(r) \\
		& =  \frac{ \Gamma(1+\alpha_{a_{ij}})  } { \Gamma(\alpha_{a_{ij}})  } \frac{\Gamma\left( \sum_{k=0}^{r}\alpha_k \right)}{\Gamma\left( 1+\sum_{k=0}^{r}\alpha_k \right)}, \qquad l= 0,\ldots, r
	\end{align*}
	where 
	$$ B(\alpha)= \frac{\prod_{k=0}^r  \Gamma(\alpha_k)  }{   \Gamma\left( \sum_{k=0}^r \alpha_k \right)} $$
	The prior for a graph $G(A)$ then becomes: 
	$$
	p\big((a_{ij})_{i<j} \mid \alpha\big) \propto \mathbb{I}_{\mathcal{C}}(G(A))\Big\{\prod_{i<j}\mathbb{P}(a_{ij}|\alpha)  \Big\}.
	$$
	Note that the prior does not integrate to 1, but has a finite mass.
	
	We assume that given $G(A)$, $\Omega$ has a $G-$Wishart prior distribution (\cite{lenk}) of parameters $\delta,D$, and given $A$  the 
	non-zero entries of $B$  are i.i.d.  univariate Gaussian random variables with mean  $\xi$ and variance $\kappa$.  
	Let $\mathcal{P}$ be the cone of $p\times p$ real-valued positive definite matrices. Then the joint prior for $\Omega$ and $B$ is given by:
	$$
	p(\Omega,B|A) = \mathbb{I}_{\mathcal{P}}(\Sigma(B,\Omega)) p(\Omega|A) p(B|A)
	$$
	which does not integrate to 1. 
	
	\subsection{Posterior Inference}
	\label{sec:alg} 
	The posterior distribution $\pi$  of all the unknown of interest is given by:
	$$
	\pi(B,\Omega,(a_{ij})_{i<j}\mid y_{1:m},\alpha) \propto p(y_{1:m}|B,\Omega,A) p(\Omega,B|A)p\big((a_{ij})_{i<j}|\alpha\big).
	$$
	The structure of the model bears some resemblance to the models considered in \cite{silva1,wang}. Besides the graph structure being different (namely, we do not allow
	bi-directional edges), we also impose the AMP constraint, mathematically represented by the term $ \mathbb{I}_{\mathcal{P}}(\Sigma(B,\Omega))$ in $p(\Omega,B|A)$, which leads to a much more
	complex structure of the graph space than the aforementioned references. As a result, posterior exploration of graph space is more challenging and computationally demanding and  requires carefully devised algorithms, in a sense more sophisticated, than the ones  considered in \cite{silva1,wang}.

	Our approach is to design an adaptive SMC sampler as in \cite{jasra} (see \cite{delm:06} for the original algorithm and \cite{beskos} for convergence results). 
	SMC based algorithms offer the advantage of being easily parallelisable, often reducing computation times over serial methods in some scenarios. The strategy involves simulating $N$ samples (particles)
	to approximate the sequence of densities
	$$
	\pi_t(B,\Omega,(a_{ij})_{i<j}\mid y_{1:m},\alpha) \propto \nu_t(B,\Omega,(a_{ij})_{i<j}|\alpha)
	$$
	where 
	$$
	\nu_t(B,\Omega,(a_{ij})_{i<j}\mid \alpha) = \Big[p(y_{1:m}\mid B,\Omega,A)\Big]^{\phi_t}p(\Omega,B\mid A)p\big((a_{ij})_{i<j}\mid \alpha\big)
	$$
	and $0=\phi_0<\cdots<\phi_T=1$. The motivation for this algorithm is well-documented  (see the aforementioned references for details), and SMC algorithms have been successfully employed in may contexts to sample from high dimensional posterior distributions. 
	
	In the implementation of the algorithm, we require a Markov kernel (e.g., a MCMC kernel) $K_t$, $t\in\{1,2,\dots,T\}$ that admits $\pi_t$ as an invariant distribution; this step is detailed in the appendix.
	The algorithm is summarised in Algorithm \ref{algo:smc_samp}. For notational convenience, we set $u_t=(B_t,\Omega_t,(a_{t,ij})_{i<j})$.
	In the initialization stage, one simulates exactly from $\pi_0$ using rejection sampling.
	The sequence of $\{\phi_t\} $ is set as proposed in \cite{Zhou 2016}, and the choice of  parameters for the MCMC kernel follows \cite{jasra}.
	
	\begin{algorithm}
		\begin{itemize}
			\item{\textbf{Initialize}. Set $t=0$, for $i\in\{1,\dots,N\}$ sample $u_0^{(i)}$ from $\pi_0$.}
			\item{\textbf{Iterate}: Set $t=t+1$. Determine $\phi_t$. If $\phi_t=1$ set  $t=T$, otherwise determine the parameters of the MCMC kernel $K_t$.
				\begin{itemize}
					\item{Resample
						$(\hat{u}_{t-1}^{(1)},\dots,\hat{u}_{t-1}^{(N)})$ using the weights $(w_{t}^{(1)},\dots,w_{t}^{(N)})$
						where, for $i\in\{1,\dots,N\}$, 
						$$
						w_t^{(i)} =  \Big(\frac{\nu_t(u_{t-1}^{(i)})}{\nu_{t-1}(u_{t-1}^{(i)})}\Big)\Big(\sum_{j=1}^N \frac{\nu_t(u_{t-1}^{(j)})}{\nu_{t-1}(u_{t-1}^{(j)})}\Big)^{-1}
						%\left\{\begin{array}{ll}
						%\frac{1}{N} %\Big(\frac{\nu_t(u_{t-1}^{(i)})q(u_{t-1}^{(i)}))}{\nu_{t-1}(u_{t-1}^{(i)})}\Big)\Big(\sum_{j=1}^N \frac{\nu_t(u_{t-1}^{(j)})q(u_{t-1}^{(j)})}{\nu_{t-1}(u_{t-1}^{(j)})}\Big)^{-1}
						% & \textrm{if}~t=1 \\
						% \Big(\frac{\nu_t(u_{t-1}^{(i)})}{\nu_{t-1}(u_{t-1}^{(i)})}\Big)\Big(\sum_{j=1}^N \frac{\nu_t(u_{t-1}^{(j)})}{\nu_{t-1}(u_{t-1}^{(j)})}\Big)^{-1} & \textrm{otherwise}.
						%\end{array}\right.
						$$%and resample  .
					}
					\item{Sample $u_t^{(i)}\mid \hat{u}_{t-1}^{(i)}$ from $K_t$ for $i\in\{1,\dots,N\}$.}
				\end{itemize}
			}
		\end{itemize}
		\caption{SMC Sampler. %We assume access to an initial $\phi_0$ and parameters of the MCMC kernel $K$
		}
		\label{algo:smc_samp}
	\end{algorithm}
	
	\section{Simulation and Real Data Study}\label{sec:res}
	In the following numerical experiments, we set $\delta=3$ and $D=I_{p}$ for the G-Wishart prior, and $\xi=0, \kappa=1$ for the distribution of the non-zero element of $B$. The number of particles for the SMC is $N=500$. The MCMC steps are Metropolis-Hastings kernels.
	
	\subsection{Simulated Example}
	
	In the simulated example, we assume that the $p$ random variables corresponding to the nodes of the graph are independent, i.e., $a_{ij} = 0 $ for all $1 \leq i,j \leq p$.
	This is a benchmark example to evaluate the ability of our algorithm to recover the structure as this is a very  simple graph structure. 
	We consider $p=10$ vertices and $m=100$ observations. To generate the data, we set  the precision matrix $\Omega$ equal to identity matrix $I_p$, and the entries of matrix $B$ are set to be $0$. For the Dirichlet prior, we consider $\alpha=(3,1,1,1)$. This prior implies a higher probability of no connection and assumes the same prior probability for any type of edge between two nodes. The reason we prefer this choice of hyper-parameters to  $\alpha=(1,1,1,1)$ (which corresponds  to a uniform prior) is due to the  computational time required by the initialization step to generate the chain graph.

	\begin{figure}[H]
		\centering\includegraphics[width=\textwidth]{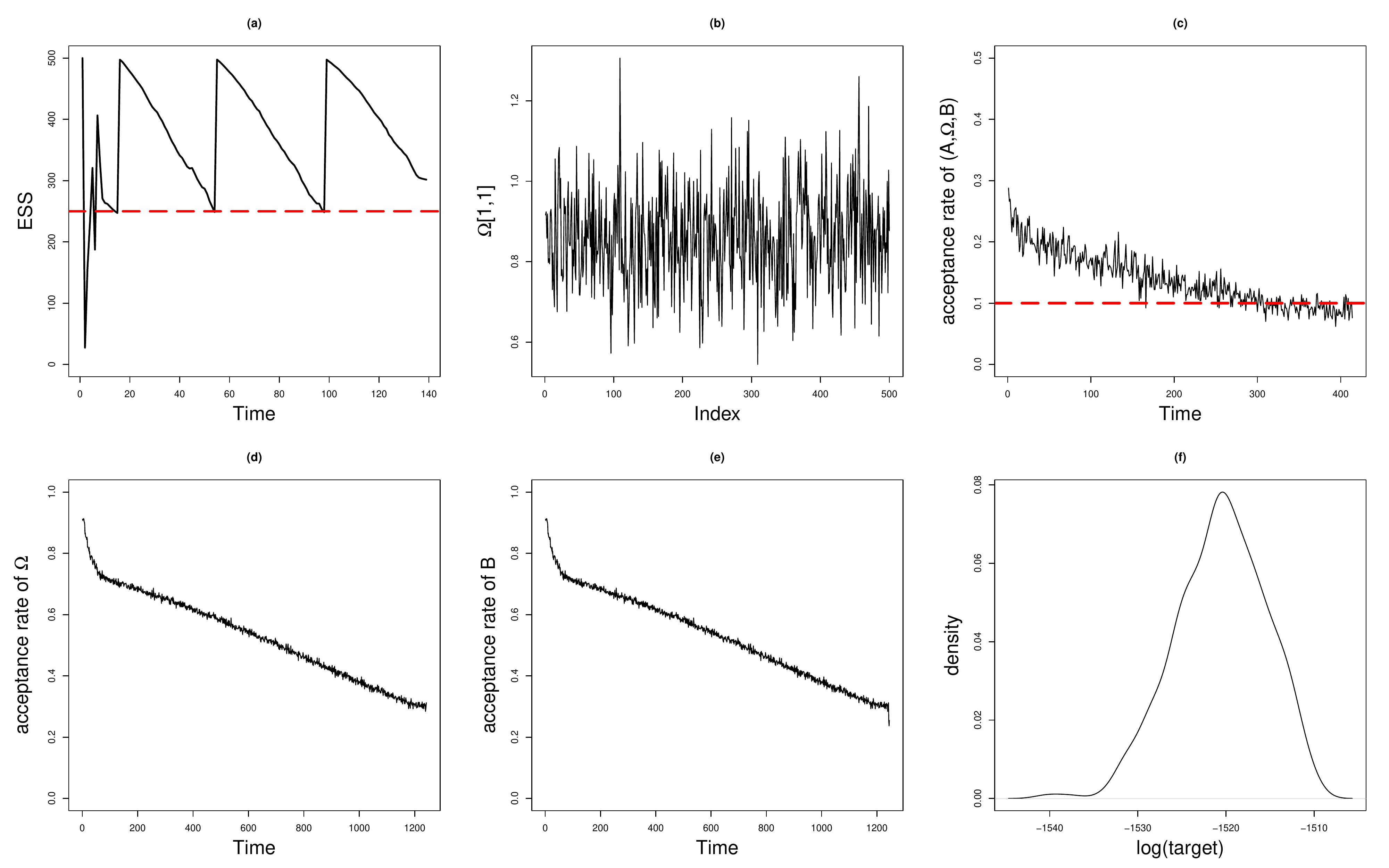}
		\caption{\textit{Simulation results for the independent case: (a) ESS in each SMC step; (b) plot of $\Omega[1,1]$ across on particles; (c) - (e) acceptance rates in each MCMC step; (f) distribution of the log(target) (i.e., log of posterior probabilities) of the samples at the end of the algorithm.}}
		\label{fig1}
	\end{figure}
	
	The simulation results are summarized in Figure \ref{fig1}. The effective sample size  (ESS) drops very fast as the algorithm begins and goes into a steady lower state after several resampling procedures. The acceptance rate is acceptable as it does not fall below a very small value.  Using the weighted samples $\{ W_{T}^{(n)} , A_{T}^{(n)}\}_{n=1}^N$ obtained at the last step of SMC, we estimate the posterior probability $\mathbb{P} (a_{ij}=0 \mid y_{1:m},\alpha), 1\leq i < j \leq p $ by
	$$  
	\widehat{\mathbb{P}} (a_{ij}=0 \mid y_{1:m},\alpha) = \sum_{p=1}^{N} W_{T}^{(p)}\mathbb{I}_{\{0\}}( a_{T,ij}^{(p)}), 
	$$
	and summarize the results in Table~1. From the table, we can see that all the estimated posterior probabilities are greater than 0.7, and most of them are above 0.9. This suggests that our algorithm is able to recover the structure of the graph (independence) used to generate the data.  
	\begin{table}[!htbp]\centering
		\caption{Posterior probability $\mathbb{P} (a_{ij}=0 | y_{1:m},\alpha), 1\leq i < j \leq p $.}
		\label{table1}
		\begin{tabular}{crrrrrrrrc}
			\hline
			Nodes &  \multicolumn{1}{c}{$2$} & \multicolumn{1}{c}{$3$} & \multicolumn{1}{c}{$4$} &
			\multicolumn{1}{c}{$5$} & \multicolumn{1}{c}{$6$} & \multicolumn{1}{c}{$7$} & \multicolumn{1}{c}{$8$} &
			\multicolumn{1}{c}{$9$} & \multicolumn{1}{c}{$10$} \\
			\hline
			$~~1$ & 0.882 & 0.892 & 0.920 & 0.932 & 0.920 & 0.870 & 0.874 & 0.908 & 0.934 \\
			$~~2$ & & 0.902 & 0.906 & 0.828 & 0.898 & 0.934 & 0.784 & 0.804 & 0.906 \\
			$~~3$ & & & 0.914 & 0.890 & 0.880 & 0.890 & 0.908 & 0.900 & 0.890 \\
			$~~4$ & & & & 0.900 & 0.866 & 0.882 & 0.770 & 0.918 & 0.900 \\
			$~~5$ & & & & & 0.788 & 0.952 & 0.912 & 0.830 & 0.918 \\
			$~~6$ & & & & & & 0.806 & 0.932 & 0.906 & 0.908 \\
			$~~7$ & & & & & & & 0.904 & 0.738 & 0.916 \\
			$~~8$ & & & & & & & & 0.918 & 0.740 \\
			$~~9$ & & & & & & & & & 0.952 \\
			\hline
		\end{tabular}
	\end{table}

	\subsection{University Graduation Rates}
	
	We investigate the performance of our algorithm when the latent graph has a more complex structure. We consider  data first presented in \cite{Druzdzel} that stem from a study for college ranking carried out in 1993. After initial analysis, Druzdzel and Glymour focus on $p=8$ variables: 
	\begin{table}[!htbp]\centering
		\begin{tabular}{ll}
			$spend$ & average spending per student,\\
			$strat$ & student-teacher ratio, \\
			$salar$ & faculty salary, \\
			$rejr$  & rejection rate, \\
			$pacc$  & percentage of admitted students who accept university's offer, \\
			$tstsc$ & average test scores of incoming students, \\
			$top10$ & class standing of incoming freshmen, and \\
			$apgra$ & average percentage of graduation.
		\end{tabular}
	\end{table}
	
\begin{figure}[!htbp]
		\begin{minipage}[b]{.5\linewidth}
			\centering
			\includegraphics[width=7cm]{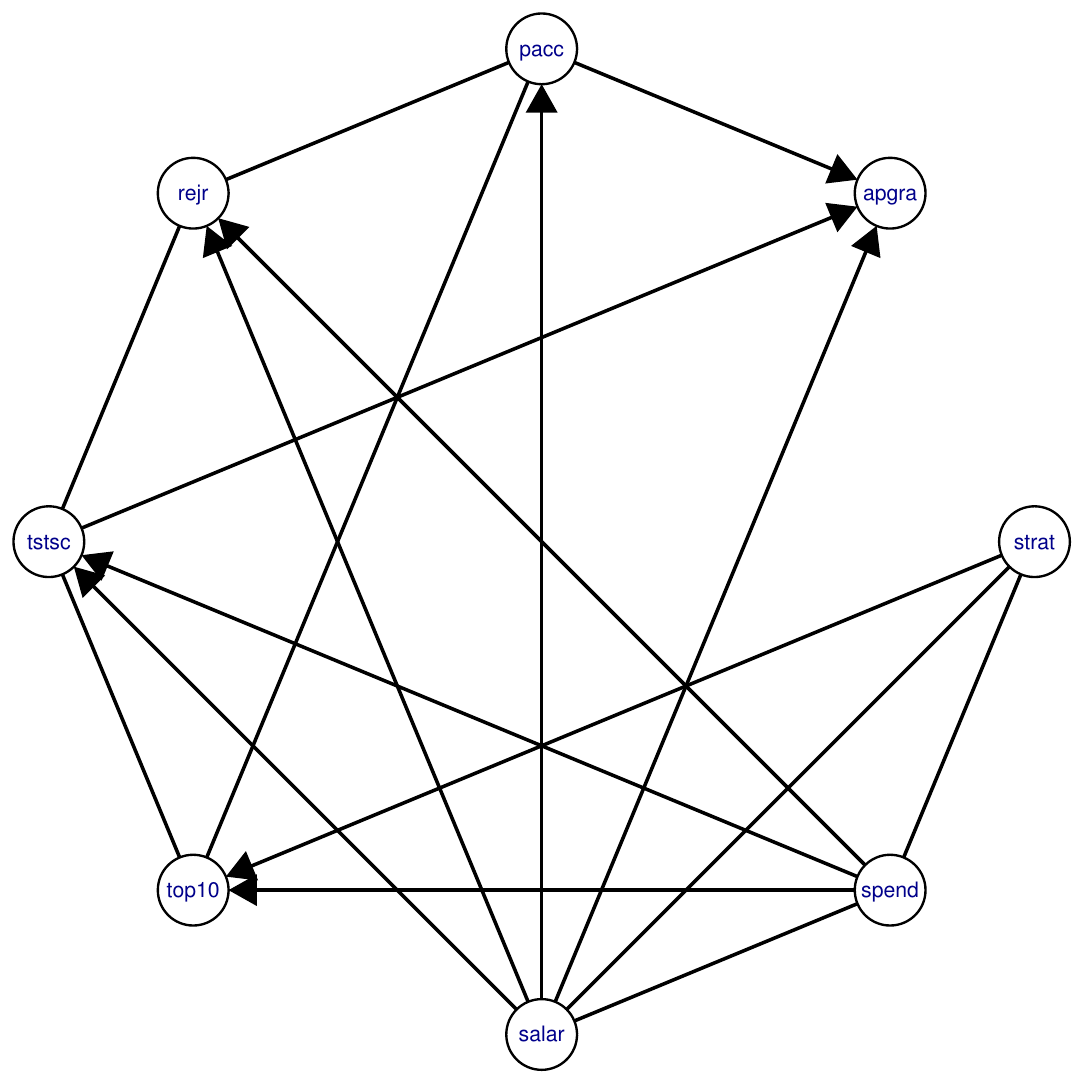}
			\caption{\textit{Chain Graph Estimate presented in \cite{drton}. }}
			\label{figSIN}
		\end{minipage}%
		\begin{minipage}[b]{.5\linewidth}
			\centering
			\renewcommand\arraystretch{1.5}
			\small
			\begin{tabularx}{\textwidth}{XYYYYYYYY}
				\hline
				& strat & spend & salar &  top10 & tstsc & rejr & pacc & apgra \\
				\hline
				strat & 0 & 1 & 1 & 2 & 0 & 0 & 0 & 0  \\
				spend & 1 & 0 & 1 & 2 & 2 & 2 & 0 & 0 \\
				salar & 1 & 1 & 0 & 0 & 2 & 2 & 2 & 2 \\
				top10 & 3 & 3 & 0 & 0 & 1 & 0 & 1 & 0 \\
				tstsc & 0 & 3 & 3 & 1 & 0 & 1 & 0 & 2 \\
				rejr & 0 & 3 & 3 & 0 & 1 & 0 & 1 & 0 \\
				pacc & 0 & 0 & 3 & 1 & 0 & 1 & 0 & 2 \\
				apgra & 0 & 0 & 3 & 0 & 3 & 0 & 3 & 0 \\
				\hline
			\end{tabularx}
			\captionof{table}{The adjacency matrix corresponding to chain graph in Figure~\ref{figSIN}.}
			\label{tableSIN}
		\end{minipage}
	\end{figure}

	Based on $m=159$ universities, the correlation matrix of these eight variables is estimated in \cite{Druzdzel}. Conditional on this correlation matrix, \cite{drton} obtain a chain graph through the SIN model selection with significance level 0.15. To get a chain graph with different AMP and LWF Markov properties, \cite{drton} further deleted the undirected edge between $top10$ and $rejr$ and the undirected edge between $salar$ and $top10$, and introduced an undirected edge between $top10$ and $rejr$. The resulting graph is shown in Figure~\ref{figSIN}, and the corresponding adjacency matrix is shown in Table~\ref{tableSIN}. This graph has three chain components $\tau_1=\{ spend, strat, salar \}, \tau_2=\{top10, tstsc, rejr, pacc \}$ and $\tau_3=\{ apgra \}$. In the original article \cite{Druzdzel} provide some insight about the causal relationship between some of the variables. The average spending per student ($spend$), student-teacher ratio ($strat$) and faculty salary ($salar$) are determined based on budget considerations and are not influenced by any of the five remaining variables. Rejection rate ($rejr$) and percentage of  students who accept the university's offer from among those who are offered admission ($pacc$) precede the average test scores ($tstsc$) and class standing ($top10$) of incoming freshmen. The latter two are measures taken over matriculating students. Finally, graduation rate ($apgra$) does not influence any of the other variables. 
	
\begin{figure}[!htbp]
		\begin{minipage}[b]{.5\linewidth}
			\centering
			\includegraphics[width=7cm]{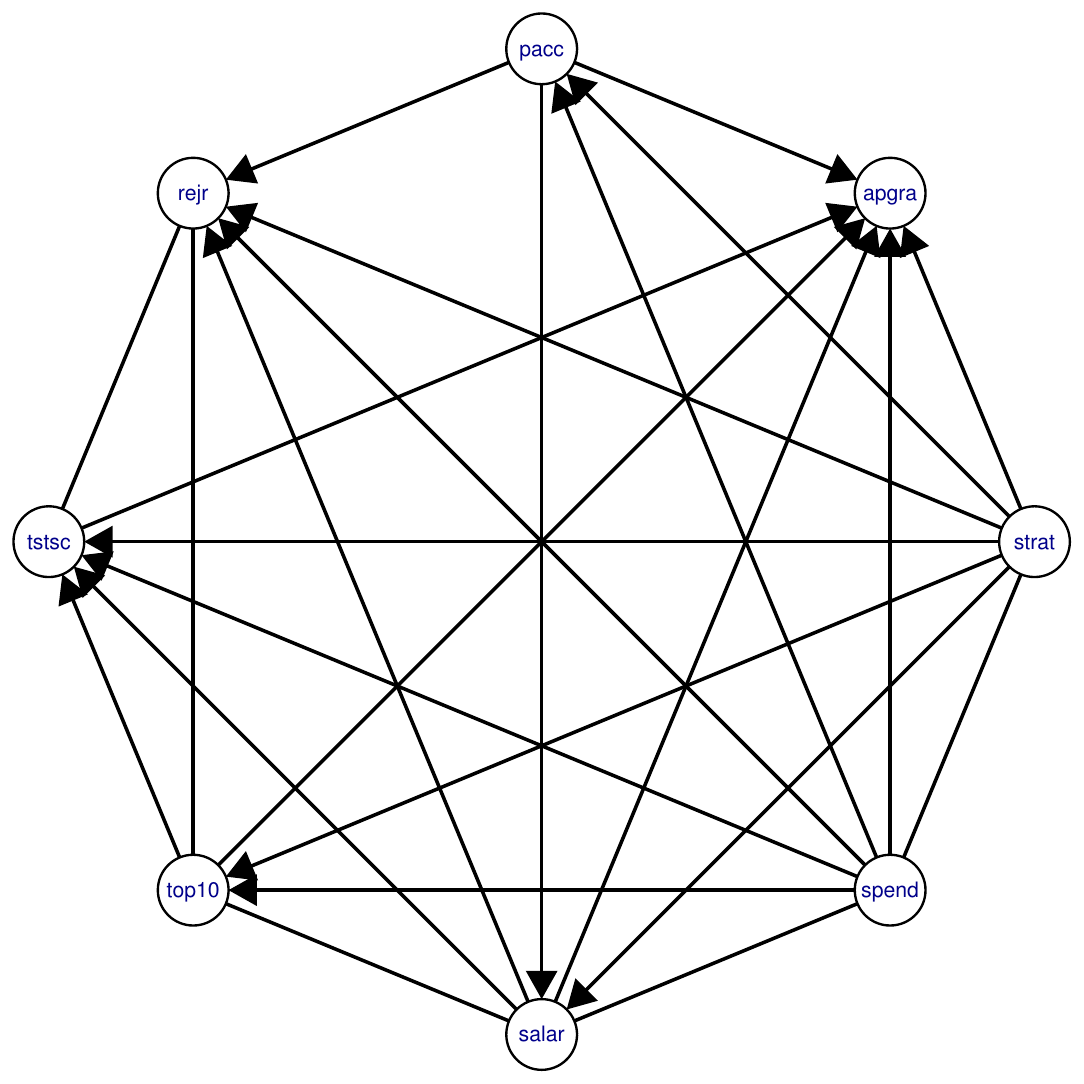}
			\caption{\textit{Empirical Graph.}}
			\label{figbase}
		\end{minipage}%
		\begin{minipage}[b]{.5\linewidth}
			\centering
			\renewcommand\arraystretch{1.5}
			\small
			\begin{tabularx}{\textwidth}{XYYYYYYYY}
				\hline
				& strat & spend & salar &  top10 & tstsc & rejr & pacc & apgra \\
				\hline
				strat & 0 & 1 & 2 & 2 & 2 & 2 & 2 & 2  \\
				spend & 1 & 0 & 1 & 2 & 2 & 2 & 2 & 2 \\
				salar & 3 & 1 & 0 & 1 & 2 & 2 & 3 & 2 \\
				top10 & 3 & 3 & 1 & 0 & 2 & 1 & 0 & 2 \\
				tstsc & 3 & 3 & 3 & 3 & 0 & 1 & 0 & 2 \\
				rejr & 3 & 3 & 3 & 1 & 1 & 0 & 3 & 0 \\
				pacc & 3 & 3 & 2 & 0 & 0 & 2 & 0 & 2 \\
				apgra & 3 & 3 & 3 & 3 & 3 & 0 & 3 & 0 \\
				\hline
			\end{tabularx}
			\captionof{table}{The adjacency matrix corresponding to the chain graph in Figure~\ref{figbase}.}
			\label{tablebase}
		\end{minipage}
	\end{figure}

	To test the performance of the proposed Bayesian chain graph model, we build an empirical baseline graph through the following procedure. Starting with a graph with only one node denoted by $G$, each time we add a new node to the existing graph. For every node in the subgraph not containing the new node, we consider the candidate graphs that are generated by adding one of four types of correlation (no edge, undirected edge and directed edges in either direction) between the new node and the node in the subgraph. We choose the graph with smallest BIC value obtained by fitting a structural equation model (SEM). SEM is a multivariate statistical analysis technique that is commonly used to analyse structural relationships.  In general, the formulation of the SEM is given by the basic equation $v=Av+u$, where $v$ and $u$ are vectors of random variables. The parameter matrix $A$ contains regression coefficients and the matrix $P=\mathbb{E}(uu^{\prime})$ gives covariances among the elements of $u$. The chain graph model described in this work can be written as $Y = BY+U$, where $U$ is a multivariate Gaussian random vector  with covariance matrix $\Omega^{-1}$. This is consistent with the structural equation model formulation. Figure~\ref{figbase} shows the derived graph and Table~\ref{tablebase} shows the corresponding adjacency matrix.
	
	We compare posterior inference from our model with the chain graphs in Figure~\ref{figSIN} and Figure~\ref{figbase}. For the SMC sampler, we set the number of samples $N=5000$. For the Dirichlet prior, we first consider a prior based on the analysis of \cite{drton}, i.e.,  choosing $\alpha=(0.39 , 0.25 , 0.36 , 0.05)$ by matching the probabilities of each type of edges in Figure~\ref{figSIN}. More precisely, the number of no-edges, un-directed edges and directed edges from $i$ to $j$ are 11, 7 and 10 according to the graphs in Figure~\ref{figSIN}. So the corresponding proportions of these three types of edge are the first three components of $\alpha$. The last component of $\alpha$ is chosen to be 0.05 to ensure the occurrence of directed edge from $j$ to $i$ and the probability of this type of edge is not large compared to the  other three types. We also perform posterior inference with $\alpha=(1,1,1,1)$, which is a uniform prior, and with $\alpha=(1,3,3,3)$, which favours more connections. The remaining parameters are specified as in the previous section. 
	
	Based on the weighted samples $\{ W_{T}^{(n)} , A_{T}^{(n)}\}_{n=1}^N$, we first estimate the posterior probability of occurrence of each edge $\widehat{\mathbb{P}} (a_{ij}=k \mid y_{1:m},\alpha),$ $1 \leq i < j \leq p $ by
	$$ 
	\widehat{\mathbb{P}} (a_{ij}=k \mid y_{1:m},\alpha) = \sum_{p=1}^{N} W_{T}^{(p)}\mathbb{I}_{\{k\}}( a_{T,ij}^{(p)}), \quad k = 0,1,2,3. 
	$$
	Then the entries of  the estimated adjacency matrix $A$ are given by 
	$$
	a_{ij} = \mathop{\arg\max}_{k=0,1,2,3} { \widehat{\mathbb{P}} (a_{ij}=k \mid y_{1:m},\alpha) }.
	$$
	
\begin{figure}[!htbp]
		\centering
		\subfigure[$\alpha=(0.39 , 0.25 , 0.36 , 0.05)$]{
			\includegraphics[width=7cm]{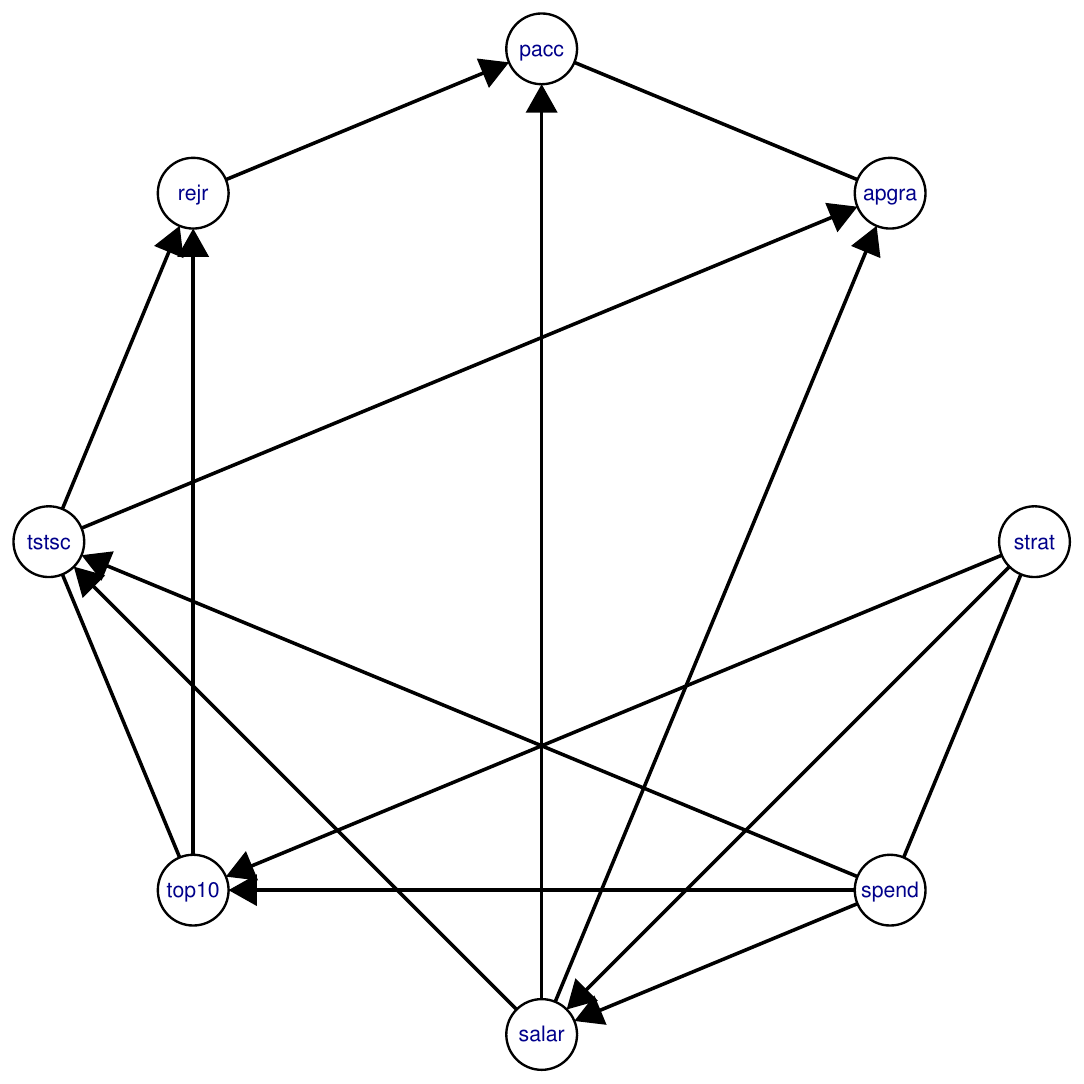}
		}
		\quad
		\subfigure[$\alpha=(1 , 1 , 1 , 1)$]{
			\includegraphics[width=7cm]{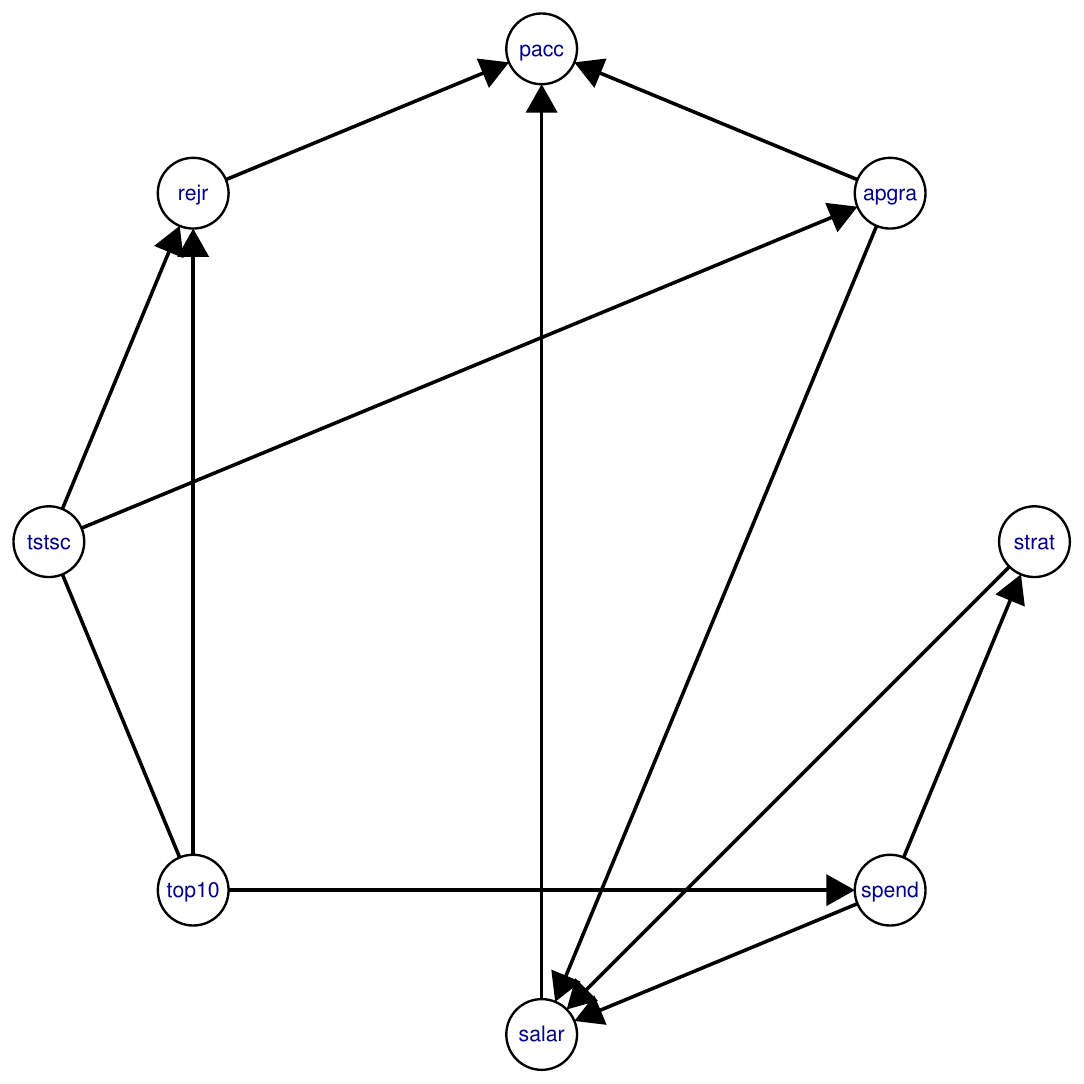}
		}
		\quad
		\subfigure[$\alpha=(1,3,3,3)$]{
			\includegraphics[width=7cm]{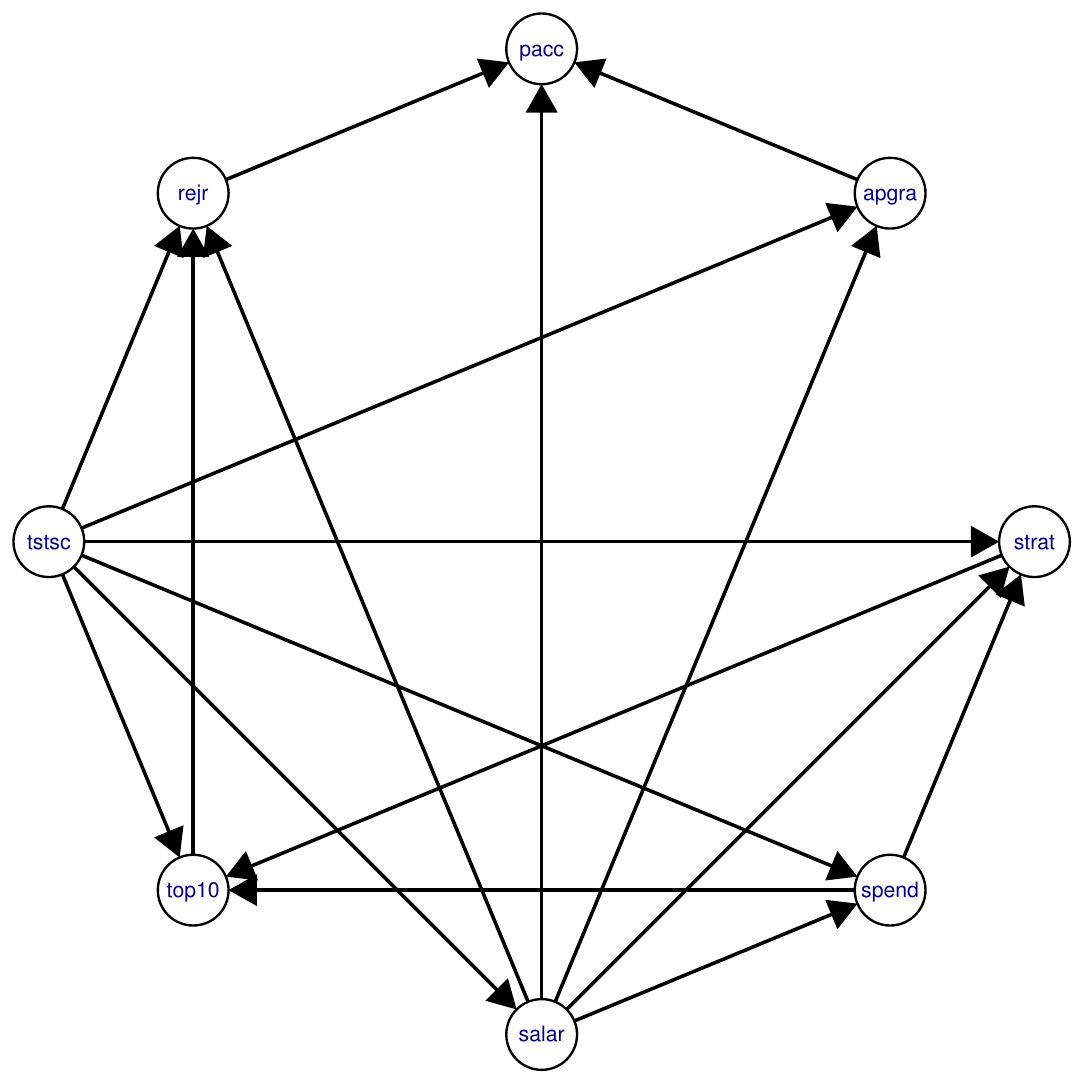}
		}
		\caption{\textit{(a): posterior estimated chain graph using a  Dirichlet prior  with $\alpha=(0.39 , 0.25 , 0.36 , 0.05)$; (b) posterior estimated chain graph using a  Dirichlet prior  with $\alpha=(1 , 1 , 1 , 1)$; (c) posterior estimated chain graph using a  Dirichlet prior  with $\alpha=(1 , 3 , 3 , 3)$.}}\label{fig_post}
	\end{figure}

	Figure \ref{fig_post} shows the estimated chain graphs obtained under each prior. The structure of the chain graph obtained from setting $\alpha=(0.39,0.25,0.36,0.05)$ is obviously more similar to Figure~\ref{figSIN} than that obtained using $\alpha=(1,1,1,1)$. This is not surprising since the prior with $\alpha=(0.39,0.25,0.36,0.05)$ is very informative and forces the posterior distribution of the edges to be similar to Figure~\ref{figSIN}. The estimated chain graph obtained setting the hyper-parameters equal to $\alpha=(1,3,3,3)$ favours more connections compared with other two priors.  If we compare our results with the conclusions in \cite{Druzdzel}, the estimated chain graph obtained under  the informative prior indeed shows that $spend$, $strat$ and $salar$ are\textit{ parents} of  other variables, and $apgra$ is   a \textit{child}  of the others. However, it does not show that $rejr$ and $pacc$ precede $tstsc$ and $top10$ (the graph just shows the opposite relation). Similarly, the graph obtained under the prior with $\alpha=(1,3,3,3)$ shows that $tstsc$ is a parent of $salar$, $strat$ and $spend$, which is a contrast to the available prior information. This can be improved by modifying the ordering of nodes and choosing a small value of $\alpha_3$.
	
	Finally, we compare our results with those obtained using SEM. We use  package {\tt sem} in R  to fit a SEM to these data. Note that when fitting the SEM model the graph topology is fixed.
	Table \ref{tablesummary} presents  a brief summary of three different chain graphs using the \textit{sem} function. From the table, we can see that the chain graph selected by the Bayesian chain graph model with prior $\alpha=(0.39,0.25,0.36,0.05)$ has smaller AIC and BIC values compared with the graph selected by the SIN model selection procedure and the empirical graph, which suggests that our algorithm can take advantage of available prior information and perform better.
	\begin{table}[!htbp]\centering
		\caption{Summaries of different chain graphs using package SEM.}
		\label{tablesummary}
		\begin{tabularx}{1\textwidth}{XY|XY|XY}
			\hline
			\multicolumn{4}{c|}{\multirow{3}*{Base chain graph}}  & \multicolumn{2}{c}{\multirow{3}*{Chain graph selected by SIN } } \\ 
			\multicolumn{4}{c|}{}&\multicolumn{2}{c}{} \\
			\multicolumn{4}{c|}{}&\multicolumn{2}{c}{} \\
			\hline
			Edge & p-value & Edge & p-value & Edge & p-value \\
			\hline
			strat --- spend & 1.630e-14 & pacc $\rightarrow$ salar & 1.137e-06 & strat --- spend & 1.630e-14 \\
			strat $\rightarrow$ salar & 1.382e-06 & pacc $\rightarrow$ rejr & 1.470e-03 & strat --- salar & 1.082e-05 \\
			spend --- salar & 2.629e-11 & strat $\rightarrow$ apgra & 8.237e-02 & strat $\rightarrow$ top10 & 1.935e-09  \\
			strat $\rightarrow$ top10 & 4.743e-07 & spend $\rightarrow$ apgra & 6.067e-02 & spend --- salar & 7.156e-13  \\
			spend $\rightarrow$ top10 & 2.822e-28 & salar $\rightarrow$ apgra & 4.794e-03 & spend $\rightarrow$ top10 & 5.979e-34  \\
			top10 --- salar & 8.931e-03 & top10 $\rightarrow$ apgra & 4.253e-01 & spend $\rightarrow$ tstsc & 3.995e-12  \\
			strat $\rightarrow$ tstsc & 3.140e-03 & tstsc $\rightarrow$ apgra & 1.096e-10 & spend $\rightarrow$ rejr & 2.909e-03  \\
			spend $\rightarrow$ tstsc & 6.634e-01 & pacc $\rightarrow$ apgra & 2.711e-03 & salar $\rightarrow$ tstsc & 2.350e-05  \\
			salar $\rightarrow$ tstsc & 2.008e-04 & & & salar $\rightarrow$ rejr & 1.323e-03  \\
			top10 $\rightarrow$ tstsc & 6.831e-19 & & & salar $\rightarrow$ pacc & 1.827e-14  \\
			strat $\rightarrow$ rejr & 1.954e-01 & & & salar $\rightarrow$ apgra & 1.570e-02  \\
			spend $\rightarrow$ rejr & 3.621e-03 & & & top10 --- tstsc & 1.256e-09  \\
			salar $\rightarrow$ rejr & 2.575e-04 & & & top10 --- pacc & 5.020e-01  \\
			top10 --- rejr & 1.816e-04 & & & tstsc --- rejr & 8.297e-03  \\
			tstsc --- rejr & 1.003e-02 & & & tstsc $\rightarrow$ apgra & 8.352e-19  \\
			strat $\rightarrow$ pacc & 2.585e-02 & & & rejr --- pacc & 5.617e-03   \\
			spend $\rightarrow$ pacc & 4.109e-07 & & & pacc $\rightarrow$ apgra & 5.481e-03  \\
			\hline
			\multicolumn{2}{c}{AIC} & \multicolumn{2}{c|}{BIC} &\centering{AIC} & BIC \\
			\multicolumn{2}{c}{67.887} & \multicolumn{2}{c|}{-13.319} & \centering{80.838} & -24.919 \\
			\hline
			\multicolumn{2}{c}{\multirow{3}*{Chain graph selected by algorithm } } \vline & \multicolumn{2}{c|}{\multirow{3}*{Chain graph selected by algorithm } } &\multicolumn{2}{c}{\multirow{3}*{Chain graph selected by algorithm } } \\ 
			&&&&&\\
			\multicolumn{2}{c}{($\alpha=(0.39 , 0.25 , 0.36 , 0.05)$)} \vline& \multicolumn{2}{c|}{($\alpha=(1,1,1,1)$)} &\multicolumn{2}{c}{($\alpha=(1,3,3,3)$)}\\
			\hline
			Edge & p-value & Edge & p-value & Edge & p-value \\
			strat --- spend & 1.630e-14 & spend $\rightarrow$ strat & 2.150e-52 & spend $\rightarrow$ strat & 1.536e-42\\
			strat $\rightarrow$ salar & 1.727e-06 & strat $\rightarrow$ salar & 9.127e-07 & salar $\rightarrow$ strat & 4.952e-06 \\
			strat $\rightarrow$ top10 & 3.597e-09 & spend $\rightarrow$ salar & 2.484e-24 & strat $\rightarrow$ top10 & 1.636e-07 \\
			spend $\rightarrow$ salar & 4.956e-33 & top10 $\rightarrow$ spend & 2.068e-25 & tstsc $\rightarrow$ strat & 8.237e-01\\
			spend $\rightarrow$ top10 & 1.086e-33 & salar $\rightarrow$ pacc & 7.304e-10 & salar $\rightarrow$ spend & 9.659e-11 \\
			spend $\rightarrow$ tstsc & 4.059e-12 & apgra $\rightarrow$ salar & 3.751e-11 & spend $\rightarrow$ top10 & 3.881e-10\\
			salar $\rightarrow$ tstsc & 5.524e-05 & top10 --- tstsc & 2.163e-14 & tstsc $\rightarrow$ spend & 2.886e-08 \\
			salar $\rightarrow$ pacc & 1.012e-18 & top10 $\rightarrow$ rejr & 2.504e-03 & tstsc $\rightarrow$ salar & 3.654e-27 \\
			salar $\rightarrow$ apgra & 3.201e-05 & tstsc $\rightarrow$ rejr & 2.847e-04 & salar $\rightarrow$ rejr & 8.844e-02 \\
			top10 --- tstsc & 2.295e-10 & tstsc $\rightarrow$ apgra & 1.428e-43 & salar $\rightarrow$ pacc & 1.062e-08 \\
			top10 $\rightarrow$ rejr & 1.951e-03 & rejr $\rightarrow$ pacc & 1.480e-04 & salar $\rightarrow$ apgra &  1.036e-04 \\
			tstsc $\rightarrow$ rejr & 2.348e-04 & apgra $\rightarrow$ pacc & 3.587e-03 & tstsc $\rightarrow$ top10 & 1.033e-20 \\
			tstsc $\rightarrow$ apgra & 1.367e-18 && & top10 $\rightarrow$ rejr & 9.160e-03 \\
			rejr $\rightarrow$ pacc & 1.032e-03 && & tstsc $\rightarrow$ rejr & 5.443e-03 \\
			pacc --- apgra & 5.315e-03 && & tstsc $\rightarrow$ apgra & 2.644e-17 \\
			&&&& rejr $\rightarrow$ pacc & 3.048e-04 \\
			&&&& apgra $\rightarrow$ pacc & 5.759e-03 \\
			\hline
			\centering{AIC} & BIC & \centering{AIC} & BIC & \centering{AIC} & BIC  \\
			\centering{61.372} & -50.522 & \centering{102.93} & -18.169 & \centering{58.401} & -47.356 \\
			\hline
		\end{tabularx}
	\end{table}

	\subsection{Tenofovir study}
	\noindent In this section, we illustrate our Bayesian  model for AMP chain graphs on a real data application from the RMP-02/MTN-006 study (\cite{AntonEtAl2012}). Tenofovir (TFV) is a medication  used to treat chronic hepatitis B and to prevent and treat HIV. TFV $1\%$ gel demonstrated $39\%$ protective efficacy in women using the gel within 12 hours before and after sexual activity in the Centre for the AIDs Programme of Research in South Africa 004 study \cite{Abdool}. Daily dosing of tenofovir disoproxil fumarate (TDF)/emtricitabine provides 62 to $73\%$ protection against HIV transmission in serodiscordant men and women enrolled in the Partners PrEP study \cite{Baeten}. The RMP-02 study was designed to evaluate the systemic safety and biologic effects of oral TFV combined with a gel formulation of the drug for application rectally and vaginally. The study enrolled 18 patients, all of whom received a single oral dose of TFV, and randomized each patient to receive either the gel formulation or a placebo several weeks later. Details about the phase 1 study are given in \cite{AntonEtAl2012}. \cite{RichardsonHarmanEtAl2014} present analyses of the ancillary studies into the treatment's biologic effects.
	
	The biologic effects we examine here concern the pharmacokinetics (PK) of TFV and its active metabolite tenofovir diphosphate (TFVdp), as well as the pharmacodynamics of the drug. The PK studies evaluate subjects' TFV and TFVdp concentrations in multiple physiologic compartments (i.e., tissues and cells) across multiple time points during the study. Table \ref{tab:Compartments} lists the compartments.
	
	\begin{table}[htp]
		\caption{Tissues and cell types examined in the PK studies}
		\begin{center}
			\begin{tabular}{llc}
				\hline
				Compound & Compartment & Notation \\
				\hline
				TFV & Blood plasma & TFV$_{plasma}$\\
				TFV & Rectal biopsy tissue & TFV$_{tissue}$\\
				TFV & Rectal fluid & TFV$_{rectal}$ \\
				TFVdp & Rectal biopsy tissue & TFVdp$_{tissue}$ \\
				TFVdp & Total mononuclear cells in rectal tissue & Total$_{\text{MMC}}$ \\
				TFVdp & CD4$^+$ lymphocytes from MMC & CD4$^+_{\text{MMC}}$ \\
				TFVdp & CD4$^-$ lymphocytes from MMC & CD4$^-_{\text{MMC}}$ \\
				\hline
			\end{tabular}
		\end{center}
		\label{tab:Compartments}
	\end{table}%
	
	\cite{RichardsonHarmanEtAl2014} demonstrate that tissue HIV infectibility (cumulative p24) is correlated with \textit{in vivo} concentrations of both TFV and TFVdp. Statistically significant, non-linear dose-response relationships with reduced tissue infectibility are found for one TFV compartment and four TFVdp compartments; the dose-response relationships are highly significant for TFVdp in whole rectal tissue, CD4$^+_{\text{MMC}}$, CD4$^-_{\text{MMC}}$ and Total$_{\text{MMC}}$ compartments. Furthermore, \cite{Yang}  conduct a comprehensive pharmacokinetic study of rectally administered TFV gel that describes the distribution of TFV and TFVdp into various tissue compartments relevant to HIV infection. They argue that TFV rectal fluid concentrations may be reasonable bio-indicators of plasma and rectal tissue concentrations, making it easier to estimate adherence and TFV concentrations in the target tissue. Therefore, the correlations between the TFV and TFVdp in compartments can be helpful in studying HIV suppression procedure and providing a measure of drug efficacy, enabling more advanced population pharmacokinetic modelling methods. In the following analysis, we investigate the correlation structure of p = 7 concentration levels in Table \ref{tab:Compartments}  collected at visit 12. From clinical knowledge, we would expect the following associations:
	\begin{itemize}
		\item TFV$_{plasma}$ is associated with TFV$_{tissue}$ (blood levels and tissue levels),
		\item TFV$_{tissue}$ is associated with TFV$_{rectal}$ (rectal tissue and rectal fluid),
		\item TFV$_{tissue}$ is associated with Total$_{\text{MMC}}$ (rectal tissue and mononuclear cells in rectal tissue),
		\item Total$_{\text{MMC}}$ is associated with CD4$^+_{\text{MMC}}$ and CD4$^-_{\text{MMC}}$ (total and constituents).
	\end{itemize}
	We normalise the observations of each variable  to have zero mean and standard deviation of one. The number of observation is $m=11$, and the number of particles is set to $N=5000$. We fit the model using as hyper-parameters in the Dirichlet prior both $\alpha=(1,1,1,1)$, which corresponds to  the uniform prior, and  $\alpha=(1,3,3,3)$, which favours the presence of connections. This latter prior choice is more suitable for small sample sizes. The remaining parameters are set as in the previous section. 
	
\begin{figure}[!htbp]
		\centering
		\subfigure[$\alpha=(1,1,1,1)$]{
			\includegraphics[width=8.3cm]{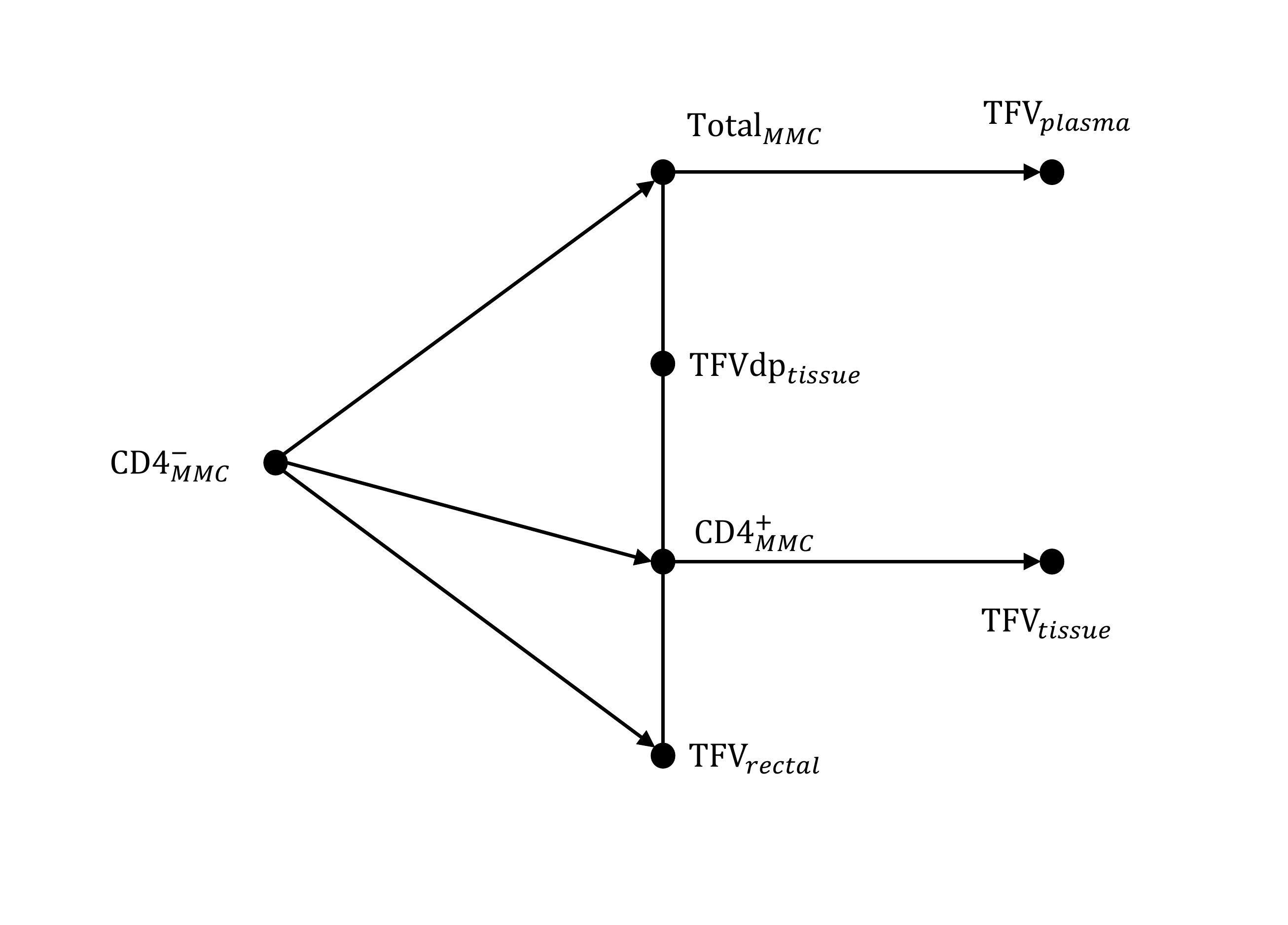}
		}
		\quad
		\subfigure[$\alpha=(1,3,3,3)$]{
			\includegraphics[width=8.3cm]{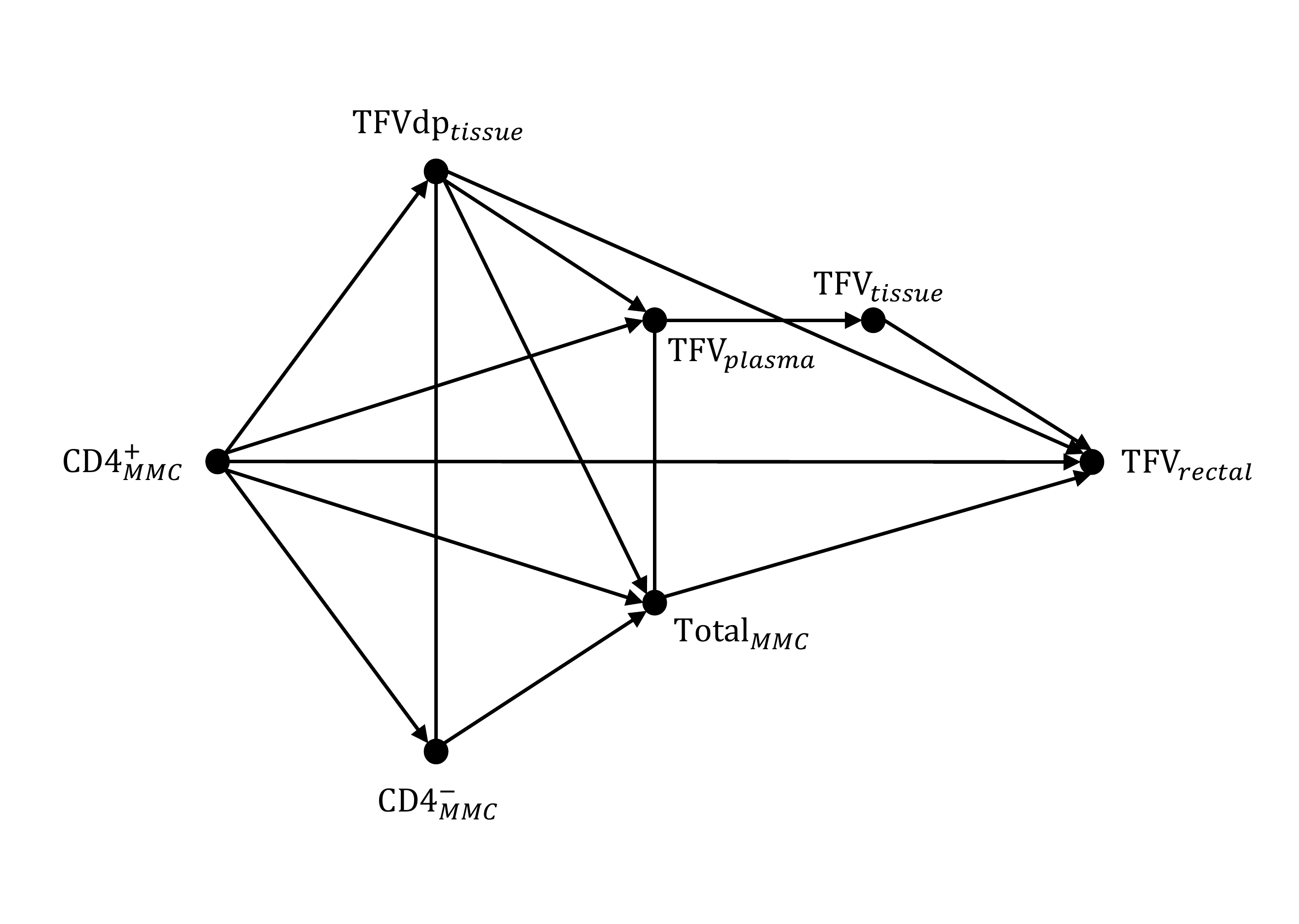}
		}
		\quad
		\caption{\textit{Chain graph with highest posterior probability.}}\label{fig.tenofovir}		
	\end{figure}

	Based on weighted samples $\{ W_{T}^{(n)} , (A,B,\Omega)_{T}^{(n)}\}_{n=1}^N$, we calculate the posterior probabilities  $\mathbb{P}\big((A,B,\Omega)_{T}^{(n)}\mid  y_{1:n},\alpha\big)$. As posterior estimate of the resulting chain graph, we report the one obtained  from the adjacency matrix $A_{T}^{(n^*)}$ where $ n^* = \mathop{\arg\max}_{n} \, \mathbb{P}\big((A,B,\Omega)_{T}^{(n)}\mid y_{1:n},\alpha\big)$, and is shown in Figure \ref{fig.tenofovir} (for both prior settings). Obviously, the chain graph obtained under the prior with $\alpha=(1,3,3,3)$ has more connections. Moreover, this graph also shows that the  TFV$_{tissue}$ is related to TFV$_{plasma}$ and TFV$_{rectal}$, the TFV$_{tissue}$ is associated with Total$_{\text{MMC}}$ through TFV$_{plasma}$, and CD4$^+_{\text{MMC}}$ causes Total$_{\text{MMC}}$ and CD4$^-_{\text{MMC}}$. This is consistent with the clinical knowledge mentioned before. However, some of these associations are missing in the chain graph obtained under the  uniform prior, e.g. the edge between TFV$_{plasma}$ and TFV$_{tissue}$ (they are independent given Total$_{\text{MMC}}$ and CD4$^+_{\text{MMC}}$ under the AMP chain graph property).
	
	We  compare our results  to those obtained by fitting a SEM model using the \textit{sem} function in the R package {\tt sem}. We summarise the results in Table \ref{tableRMP}. The AIC and BIC values of the model corresponding to the chain graph obtained under the  uniform prior are missing, which may due to the singular Hessian matrix when estimating the covariance matrix of the parameter. From the table, it is evident that the p-values of some edges are large. For example, the edge CD4$^+_{\text{MMC}}$ $\rightarrow$ TFV$_{tissue}$ in the chain graph obtained under uniform prior has a p-value 0.704. These results suggest that these relationships are not significant in a SEM model.  
	
	When inadequate fit of a structural equation model is observed, model modification is often conducted followed by retesting of the modified model. The most popular statistic is the modification index, which is a chi-square score test statistic with degree of freedom one. The modification index provides an estimated value in which the model's chi-square test statistic would decrease if the corresponding parameter is added to the model and respecified as a free parameter. We perform model modification using the \textit{modIndices} function in {\tt sem} package, which calculates modification indices and estimates parameter changes for the fixed and constrained parameters in a structural equation model. Table 5 shows the five largest modification indices for both the  $A$ matrix and $P$ matrix for the model corresponding to the chain graph obtained under the Dirichlet prior with $\alpha=(1,3,3,3)$. The modification indices suggest that a better fit to the data would be achieved by adding association between TFV$_{rectal}$ and TFV$_{plasma}$ to the model. The small sample size makes it obviously challenging to estimate associations. Furthermore, the level in rectal tissue may depend on whether or not the patient received the TFV gel or the placebo, which may make the correlation of TFV$_{tissue}$ with other variables unclear. 
	%This can be improved by introducing covariates to include the influence of TFV gel.
	
	\begin{table}[!htbp]\centering
		\caption{Summaries of two different chain graphs using package SEM.}
		\label{tableRMP}
		\begin{tabularx}{1.03\textwidth}{XY|XY}
			\hline	
			\multicolumn{2}{c|}{\multirow{3}*{Chain graph selected by algorithm}} & \multicolumn{2}{c}{\multirow{3}*{Chain graph selected by algorithm } } \\ 
			&&&\\
			\multicolumn{2}{c}{($\alpha=(1,1,1,1)$)}\vline & \multicolumn{2}{c}{($\alpha=(1,3,3,3)$)} \\
			\hline
			Edge & p-value & Edge & p-value  \\
			CD4$^-_{\text{MMC}}$ $\rightarrow$ Total$_{\text{MMC}}$ & 9.881e-19 & CD4$^+_{\text{MMC}}$ $\rightarrow$ TFVdp$_{tissue}$ & 5.687e-01 \\
			CD4$^-_{\text{MMC}}$ $\rightarrow$ CD4$^+_{\text{MMC}}$ & 4.091e-81 &
			CD4$^+_{\text{MMC}}$ $\rightarrow$ CD4$^-_{\text{MMC}}$ & 2.941e-46 \\
			CD4$^-_{\text{MMC}}$ $\rightarrow$ TFV$_{rectal}$ & 1.496e-01 & CD4$^+_{\text{MMC}}$ $\rightarrow$ TFV$_{plasma}$ & 1.210e-02 \\
			Total$_{\text{MMC}}$ --- TFVdp$_{tissue}$ & 1.589e-02 & CD4$^+_{\text{MMC}}$ $\rightarrow$ Total$_{\text{MMC}}$ & 7.028e-10 \\
			TFVdp$_{tissue}$ --- CD4$^+_{\text{MMC}}$ & 1.812e-02 & CD4$^+_{\text{MMC}}$ $\rightarrow$ TFV$_{rectal}$ & 1.874e-03 \\
			CD4$^+_{\text{MMC}}$ --- TFV$_{rectal}$ & 8.583e-03 & TFVdp$_{tissue}$ --- CD4$^-_{\text{MMC}}$ & 8.477e-02 \\
			Total$_{\text{MMC}}$ $\rightarrow$ TFV$_{plasma}$ & 1.815e-03 & TFVdp$_{tissue}$ $\rightarrow$ TFV$_{rectal}$ & 2.991e-01 \\
			CD4$^+_{\text{MMC}}$ $\rightarrow$ TFV$_{tissue}$ & 7.043e-01 & TFVdp$_{tissue}$ $\rightarrow$ TFV$_{plasma}$ & 2.584e-01 \\
			&& TFVdp$_{tissue}$ $\rightarrow$ Total$_{\text{MMC}}$ & 1.162e-13 \\
			&& CD4$^-_{\text{MMC}}$ $\rightarrow$ Total$_{\text{MMC}}$ & 2.352e-22 \\
			&&  TFV$_{plasma}$ $\rightarrow$ TFV$_{tissue}$ & 4.259e-01 \\
			&&  TFV$_{plasma}$ --- Total$_{\text{MMC}}$ & 5.719e-02 \\
			&& TFV$_{tissue}$  $\rightarrow$ TFV$_{rectal}$ & 7.550e-01 \\
			&& Total$_{\text{MMC}}$ $\rightarrow$ TFV$_{rectal}$ & 2.337e-02\\
			\hline
			\centering{AIC} & BIC & \centering{AIC} & BIC \\
			\centering{Inf} & Inf & \centering{46.875} & -11.909 \\
			\hline
		\end{tabularx}
	\end{table}
	
	\begin{table}[!htbp]\centering
		\caption{Summaries of modification indices for the model corresponding to the chain graph obtained under prior with $\alpha=(1,3,3,3)$.}
		\label{tableRMP}
		\begin{tabularx}{\textwidth}{XY|XY}
			\hline	
			\multicolumn{2}{c|}{\multirow{3}*{5 largest modification indices, $A$ matrix}} & \multicolumn{2}{c}{\multirow{3}*{5 largest modification indices, $P$ matrix} } \\ 
			&&&\\
			\multicolumn{2}{c}{(regression coefficients)}\vline & \multicolumn{2}{c}{(variances/covariances)} \\
			\hline
			TFV$_{rectal}$ $\rightarrow$ Total$_{\text{MMC}}$ & 3.281 & TFV$_{rectal}$ --- Total$_{\text{MMC}}$ & 3.394  \\
			TFV$_{rectal}$ $\rightarrow$ TFV$_{plasma}$ & 2.219 & TFV$_{rectal}$ --- TFV$_{plasma}$ & 2.881 \\
			CD4$^-_{\text{MMC}}$ $\rightarrow$ TFV$_{rectal}$ & 0.709 & TFV$_{rectal}$ --- CD4$^-_{\text{MMC}}$ & 0.709 \\
			TFV$_{rectal}$ $\rightarrow$ TFVdp$_{tissue}$ & 0.654 & TFV$_{rectal}$ --- TFVdp$_{tissue}$ & 0.709 \\
			TFV$_{rectal}$ $\rightarrow$ CD4$^-_{\text{MMC}}$ & 0.555 & TFV$_{tissue}$ --- TFVdp$_{tissue}$  & 0.389 \\
			\hline
		\end{tabularx}
	\end{table}
	
	\section{Conclusions}
	In this article we propose a novel Bayesian model for latent  AMP chain graphs, for which observations are available only on the nodes of the graph. Posterior inference is performed through a specially devised SMC algorithm. We investigate the ability of the model to recover a  range of structures, also when prior knowledge is available. The performance of the SMC sampler is stable and consistent in our numerical study. However, the sampler is not suitable when the number of nodes $p$ is large, as the initializing step is difficult. Moreover, the proposed algorithm does not scale well with respect to $p$. The computational cost of computing the probabilities in the adjacency matrix is $\mathcal{O}(p^2)$, and the computation of the normalizing constant in the $G$-Wishart distribution is quite expensive when $p$ is large (approximately  $\mathcal{O}(p^3)$). 
	
	Several extensions of this work are possible. First, the algorithm can be extended to large $p$ by choosing a more efficient initial proposal $q$ and by actually exploiting parallel computing techniques. Second, the model can be extended to accommodate multiple groups of observations, allowing borrowing information across groups or time periods. Third, we could avoid  sampling $\Omega$ and $B$ by using a Laplace approximation when calculating the posterior probability.
	
	\appendix
	
	\section{MCMC Kernel}
	
	The updates are taken in the order $A,\Omega, B$ together, then $\Omega$ and $B$. We use Metropolis-Hastings steps. The updates for $\Omega$ and $B$ are preformed element-wise.
	For each element of $\Omega$ and $B$ a Gaussian random walk with variance $\sigma_1^2$ is used as proposal. The acceptance probability is calculated as usual and note that if any of the constraints on $\Omega$ or $B$
	are violated, then  the acceptance probability is zero.
	
	The update for $A,\Omega, B$ together is much more complicated. We denote with  $(A^{(n)},\Omega^{(n)}, B^{(n)})$  the current values of $A,\Omega, B$ and the proposed states with 
	$(A_c^{(n)},\Omega_c^{(n)}, B_c^{(n)})$. The move proceeds by picking an $(a_{ij}^{(n)})_{i<j}$ uniformly at random and then proposing one of the other possible 3 values with uniform probability. Note that at this stage, we will check if the proposed graph is a chain graph and if not, the move is rejected instantly.
	Based upon this proposal, we propose new values of $\Omega$ and $B$ as detailed below. Suppose the selected edge is $(i,j), i<j$, consider the following scheme:
	\begin{enumerate}[(i)]

		\item If the transformation of the selected edge is $``0\rightarrow 1"$, we take $\Omega_{c}^{(n)}=\Omega^{(n)}$ except for elements $\Omega_{c}^{(n)}[i,j]$ and $\Omega_{c}^{(n)}[j,i]$. For these elements, we set $\Omega_{c}^{(n)}[i,j] \sim \mathcal{N}(0,\sigma^2_2)$ and $\Omega_{c}^{(n)}[j,i]=\Omega_{c}^{(n)}[i,j]$. Finally, we set $B_{c}^{(n)}=B^{(n)}$.
		
		\item If the transformation of the selected edge is $``0\rightarrow 2"$, we take $B_{c}^{(n)}=B^{(n)}$ except for element $B_{c}^{(n)}[j,i]$. For this element, we set $B_{c}^{(n)}[j,i] \sim \mathcal{N}(\xi,\kappa)$. Finally, we set $\Omega_{c}^{(n)}=\Omega^{(n)}$.
		
		\item If the transformation of the selected edge is $``0\rightarrow 3"$, we take $B_{c}^{(n)}=B^{(n)}$ except for element $B_{c}^{(n)}[i,j]$. For this element, we set $B_{c}^{(n)}[i,j] \sim \mathcal{N}(\xi,\kappa)$. Finally, we set $\Omega_{c}^{(n)}=\Omega^{(n)}$.
		
		\item If the transformation of the selected edge is $``1\rightarrow 0"$, we take $\Omega_{c}^{(n)}=\Omega^{(n)}$ except for elements $\Omega_{c}^{(n)}[i,j]$ and $\Omega_{c}^{(n)}[j,i]$. For these elements, we set $\Omega_{c}^{(n)}[i,j]=\Omega_{c}^{(n)}[j,i]=0 $. Finally, we set $B_{c}^{(n)}=B^{(n)}$.
		
		\item If the transformation of the selected edge is $``1\rightarrow 2"$, we take $\Omega_{c}^{(n)}=\Omega^{(n)}$ except for elements $\Omega_{c}^{(n)}[i,j]$ and $\Omega_{c}^{(n)}[j,i]$. For these elements, we set $\Omega_{c}^{(n)}[i,j]=\Omega_{c}^{(n)}[j,i]=0 $. Finally, we take $B_{c}^{(n)}=B^{(n)}$ except for element $B_{c}^{(n)}[j,i]$. For this element, we set $B_{c}^{(n)}[j,i] \sim \mathcal{N}(\xi,\kappa)$.
		
		\item If the transformation of the selected edge is $``1\rightarrow 3"$, we take $\Omega_{c}^{(n)}=\Omega^{(n)}$ except for elements $\Omega_{c}^{(n)}[i,j]$ and $\Omega_{c}^{(n)}[j,i]$. For these elements, we set $\Omega_{c}^{(n)}[i,j]=\Omega_{c}^{(n)}[j,i]=0 $. Finally, we take $B_{c}^{(n)}=B^{(n)}$ except for element $B_{c}^{(n)}[i,j]$. For this element, we set $B_{c}^{(n)}[i,j] \sim \mathcal{N}(\xi,\kappa)$.
		
		\item If the transformation of the selected edge is $``2\rightarrow 0"$, we take $B_{c}^{(n)}=B^{(n)}$ except for element $B_{c}^{(n)}[j,i]$. For this element, we set $B_{c}^{(n)}[j,i]=0$. Finally, we set $\Omega_{c}^{(n)}=\Omega^{(n)}$.
		
		\item If the transformation of the selected edge is $``2\rightarrow 1"$, we take $B_{c}^{(n)}=B^{(n)}$ except for element $B_{c}^{(n)}[j,i]$. For this element, we set $B_{c}^{(n)}[j,i]=0$. Finally, we take $\Omega_{c}^{(n)}=\Omega^{(n)}$ except for elements $\Omega_{c}^{(n)}[i,j]$ and $\Omega_{c}^{(n)}[j,i]$. For these elements, we set $\Omega_{c}^{(n)}[i,j] \sim \mathcal{N}(0,\sigma^2_2)$ and $\Omega_{c}^{(n)}[j,i]=\Omega_{c}^{(n)}[i,j]$.
		
		\item If the transformation of the selected edge is $``2\rightarrow 3"$, we take $B_{c}^{(n)}=B^{(n)}$ except for elements $B_{c}^{(n)}[j,i]$ and $B_{c}^{(n)}[i,j]$. For these elements, we set $B_{c}^{(n)}[j,i]=0$, and $B_{c}^{(n)}[i,j] \sim \mathcal{N}(\xi,\kappa)$. Finally, we set $\Omega_{c}^{(n)}=\Omega^{(n)}$.
		
		\item If the transformation of the selected edge is $``3\rightarrow 0"$, we take $B_{c}^{(n)}=B^{(n)}$ except for element $B_{c}^{(n)}[i,j]$. For this element, we set $B_{c}^{(n)}[i,j]=0$. Finally, we set $\Omega_{c}^{(n)}=\Omega^{(n)}$.
		
		\item If the transformation of the selected edge is $``3\rightarrow 1"$, we take $B_{c}^{(n)}=B^{(n)}$ except for element $B_{c}^{(n)}[i,j]$. For this element, we set $B_{c}^{(n)}[i,j]=0$. Finally, we take $\Omega_{c}^{(n)}=\Omega^{(n)}$ except for elements $\Omega_{c}^{(n)}[i,j]$ and $\Omega_{c}^{(n)}[j,i]$. For these elements, we set $\Omega_{c}^{(n)}[i,j] \sim \mathcal{N}(0,\sigma^2_2)$ and $\Omega_{c}^{(n)}[j,i]=\Omega_{c}^{(n)}[i,j]$.
		
		\item If the transformation of the selected edge is $``3\rightarrow 2"$, we take $B_{c}^{(n)}=B^{(n)}$ except for elements $B_{c}^{(n)}[i,j]$ and $B_{c}^{(n)}[j,i]$. For these elements, we set $B_{c}^{(n)}[i,j]=0$, and $B_{c}^{(n)}[j,i] \sim \mathcal{N}(\xi,\kappa)$. Finally, we set $\Omega_{c}^{(n)}=\Omega^{(n)}$.

	\end{enumerate}
	The acceptance probabiity for this move is easily calculated.

\end{document}